\documentclass[10pt]{iopart}
\usepackage{graphicx}
\usepackage{float}

\usepackage{multirow}
\usepackage{hyperref}
\usepackage{enumerate}
\usepackage{listings}

\usepackage{iopams}

\expandafter\let\csname equation*\endcsname\relax
\expandafter\let\csname endequation*\endcsname\relax 
\usepackage{amsmath}  

\usepackage[]{ifthen}
\newboolean{bpublic}

\setboolean{bpublic}{false} 

		\newcommand{\alice}{ALICE}
        \newcommand{\alicee}{A Large Ion Collider Experiment}
        \newcommand{\alien}{AliEn}
        \newcommand{\aliene}{ALICE Environment}
        \newcommand{\aliroot}{AliRoot}
        \newcommand{\cern}{CERN}
        \newcommand{\lhc}{LHC}
        \newcommand{\lhce}{Large Hadron Collider}
        \newcommand{\wlcg}{WLCG}
        \newcommand{\wlcge}{Worldwide LHC Computing Grid}
        \newcommand{\egee}{EGEE}
        \newcommand{\atlas}{ATLAS}
        \newcommand{\lhcb}{LHCb}

\begin{document}

\title{A Mediated Definite Delegation Model allowing for Certified Grid Job Submission}

	\author{Steffen Schreiner$^1$$^,$$^2$, Latchezar Betev$^1$, Costin
Grigoras$^1$, Maarten Litmaath$^1$}
	\address{$^1$ European Organization for Nuclear Research CERN, Geneva,
	Switzerland}
	\address{$^2$ Center for Advanced Security Research Darmstadt - CASED\\~~~and
	Technische Universit\"{a}t Darmstadt, Germany}
	\ead{steffen.schreiner@cern.ch}

\bibliographystyle{iopart-num}


\begin{abstract}
Grid computing infrastructures need to provide
traceability and accounting of their users' activity and
protection against misuse and privilege escalation.
A central aspect of multi-user Grid job environments is
the necessary delegation of privileges in the course of a job submission. 
With respect to these generic requirements
this document describes an improved handling of multi-user
Grid jobs in the \alice\ (``\alicee'') Grid Services.\\
A security analysis of the \alice\ Grid
job model is presented with derived security objectives, followed by a
discussion of existing approaches of unrestricted delegation based on X.509
proxy certificates and the Grid middleware gLExec. Unrestricted
delegation has severe security consequences and limitations,
most importantly allowing for identity theft and forgery of delegated assignments.
These limitations are discussed and formulated, both in general and with respect
to an adoption in line with multi-user Grid jobs. Based on the architecture of
the \alice\ Grid Services, a new general model of mediated definite delegation is
developed and formulated, allowing a broker to assign context-sensitive user
privileges to agents. The model provides strong accountability and long-term 
traceability. A prototype implementation allowing for certified Grid jobs is
presented including a potential interaction with gLExec. The achieved 
improvements regarding system security, malicious job exploitation, identity
protection, and accountability are emphasized, followed by a
discussion of non-repudiation in the face of malicious Grid jobs.
\end{abstract}

\section{Introduction}
Global eScience Grid infrastructures provide researchers with unified access
to computing and storage services across national borders, jurisdictions, and
domains of responsibility. Accordingly and beyond the existence of operational
and usage policies, accountability needs to be ensured for actions occurring 
due to the operation of such infrastructures and 
in the course of its users' activities. Protection against
misuse and privilege escalation needs to be established and any violations need
to be traceable. These baseline security concerns form
the general background and motivation of the work presented in this document.\\
The \alice\ (``\alicee'') Grid
Services\ifthenelse{\boolean{bpublic}}{~\cite{aliceano}}{~\cite{alice}}, a
globally distributed Storage and Computation Grid, are developed and operated by 
the \alice\
Collaboration\ifthenelse{\boolean{bpublic}}{~\cite{collabano}}{~\cite{collab}}
as a research cyberinfrastructure. Its central 
Workload Management System (WMS) and File Catalogue are provided by the open
source Grid framework \alien
(``\aliene'')\ifthenelse{\boolean{bpublic}}{~\cite{alienano,alien03ano}}{~\cite{alien,alien03}}.
The system provides the infrastructure for simulation, reconstruction and analysis of physics data collected by the \alice\
detector at \cern, one of the four large experiments within the \lhce\ (\lhc).
As such, it is embedded within the \wlcge\
(\wlcg)\ifthenelse{\boolean{bpublic}}{~\cite{wlcgano}}{~\cite{wlcg}}, a tiered
infrastructure of Grid services for the large \lhc\ experiments. The \alice\ Grid Services constitute a Virtual Organization (VO) 
and are based on 75 computing centres (hereafter
referred to as \textit{Sites}) located in 33 countries, 
combining up to 35k CPU cores and 500PB of storage, and serving approximately 1000
users within the collaboration.\\
The File Catalogue establishes a central logical layer on top
of a globally distributed set of storage servers provided by the Sites, which constitutes one
virtual Grid File System for \alice. The Computation Grid is based on Worker
Nodes (WNs) aggregated on Sites within batch farms,
receiving Grid jobs from the central WMS. The jobs
are specified and represented by a textual description called \textit{JDL} (originating from
Job Description Language), listing e.g. the job ID reference
number, the file to be executed, execution arguments, and input and output
files. Upon submission, Grid jobs are placed into a central task queue as part
of the WMS, from which they are
fetched for execution depending on order of priority and dependency matching.
WMS and File Catalogue
form together the so-called \textit{Central Services}.\\
Fundamentally, the \alice\ Grid Services provide two functionality
or use cases: the Grid File System access and the dispatch of Grid
jobs. Both are surrounded by a rich set of corresponding management and
maintenance functionality.
The general program code executed within Grid jobs is composed of centrally
provided software packages, most importantly the
\aliroot\ifthenelse{\boolean{bpublic}}{~\cite{alirootano}}{~\cite{aliroot}}
software framework, and user supplied code. Thereby, users are free to upload program code and data of 
any kind into the Grid File System and request it to be executed in Grid jobs.
They are legally obliged though, to only utilize the \alice\ Grid Services for their research within the
\alice\ experiment and as such use Grid jobs in order to analyse the experiment's
data.\\
The freedom of the system as a globally
distributed cyberinfrastructure creates a challenge for security and in
particular of accountability and liability. This document presents an in-depth
security analysis of the submission model in the \alice\ Grid Services and discusses
objectionable and even severe security 
problems. In the course of the assessment of the
applicability of gLExec and Multi-User Pilot Jobs in \alice\, we identified the
concept of unrestricted delegation based on X.509
proxy certificates as a mechanism for Grid user credentials
 to be deficient 
and the
its adoption highly questionable. In particular, we found
PCs ineligible regarding \textit{accountability}, \textit{non-repudiation},
and due to the potential occurrence of \textit{identity theft}.
This document presents the results of
this analysis and proposes an alternative approach of \textit{certified} Grid jobs.\\
In the remainder of this introduction, 
access control and accountability (section \ref{supj} ) and the
implementation of the Pilot Job concept (section \ref{pilots} ) within the \alice\
Grid Services are described. Chapter \ref{secsupj } presents a security analysis
of Single-User Pilot Jobs in the \alice\ Grid Services and 
objectives derived therefrom. Within
chapter \ref{mupj}, the concept of Multi-User Pilot Jobs with gLExec based on
proxy credentials is described, general limitations of unrestricted
delegation with proxy credentials are assessed (section \ref{proxies} ), and
the necessary propagation of of these credentials is examined
(section \ref{prop} ). This is followed by an analysis and specification of the
resulting security problems of proxy credentials based on unrestricted
delegation with respect to Multi-User Pilot Jobs (section \ref{problems} ).
A new model of delegation is presented throughout
chapter \ref{model} and an implementation of the model presented, allowing for
certified Grid jobs (section \ref{sjdl} ) and a potential interaction with gLExec
(section \ref{sjdlgl} ). The concept of non-repudiation in the face of an
occurred security incident within a certified Grid job is discussed (section
\ref{nrepud} ), followed by a review of the remaining objective of integrity of
a Grid job's environment on a WN. Finally, the new model and its implementation 
are reconsidered with respect to related work (section \ref{related} ).

\subsection{User access in the \alice\ Grid Services}
\label{supj}
Throughout the \wlcg, X.509 certificates~\cite{rfc5280} are used as the
basic mechanism for authentication and authorization of Grid users and
operators, through middleware provided e.g.\ by the Globus Toolkit~\cite{globus}, in which the
certificates are signed by \wlcg\
recognized Certificate Authorities. Such a user certificate is hereafter
referred to as a \textit{Grid Certificate}, while the term \textit{Grid
Credential} is used in order to denote the pair of a user's Grid Certificate
and the corresponding private key.\\
Based on the concept of X.509 proxy certificates~\cite{proxy,rfc3820}, derived
Proxy Credentials (PCs) are further used in order to allow for delegation and
single sign-on. Such PCs consist of a generated private/public key pair
and a Grid Certificate, whereat the public key is signed with the private
key corresponding to the Grid Certificate.\\
In the \alice\ Grid Services, these user PCs are used only upon client
logon for authentication and authorization. After a PC with its Grid
Certificate is validated in order to grant access, the Grid Certificate's
Subject entry is mapped to an \alice-internal user name based on a LDAP service
(see figure~\ref{supj}).
Throughout the whole system, Grid users are only represented by their 
\alice-internal user name and there is no actual mechanism for delegation of
privileges embedded in the system. Whenever a Grid job is submitted
by a client, the corresponding \alice\ user name is associated to the JDL within
the Central Services. Grid jobs are executed on WNs of Sites on behalf of the
\alice\ VO and user accountability is established based on the internal user name.

\subsection{Single-User Pilot Jobs and the \alien\ JobAgent}
\label{pilots}
Pilot jobs (PJs) in Grid environments
implement an approach to optimize resource utilization and in particular to
set up and assess the local environment before a Grid job execution
on a WN. The basic principle is to let a WN
start running a Grid service instead of an actual job or payload and to
let this service handle the execution of one or more actual jobs.
Due to the
concept of PJs, the actual
underlying WN and the batch system in which it is embedded can be fully
abstracted and a virtual layer is
built on top of all Sites and their batch systems.\\
In case of the \alice\ Grid Services, a Computing Element (a
Site service functioning as a resource broker for its batch system) advertises
idle resources to the Central Services. In line with these advertisements, Sites are
requested to start \textit{JobAgents} (JAs), the Pilot Job implementation of
\alien, according to the load in the central task queue and the Sites'
capabilities (see figure~\ref{supj}, red arrows I - V). Once these JAs have started,
they will call the Central Services and request actual Grid jobs. Thereby,
for each Site the 
submission of JA requests follows the overall queue status and
each Site will receive a request for each matching job
in the queue. As a consequence, 
JAs compete for jobs and the execution of jobs will be
assigned as fast as possible with respect to
the overall state of the Grid. The JA prepares a job's environment by
installing the \alice\ software specified in the job's JDL and not yet present,
subsequently advertises the corresponding 
capabilities and can execute a number of consecutive jobs
within its own specified lifetime.\\
\begin{figure}[h]
\begin{center}
\includegraphics[width=\textwidth,keepaspectratio]{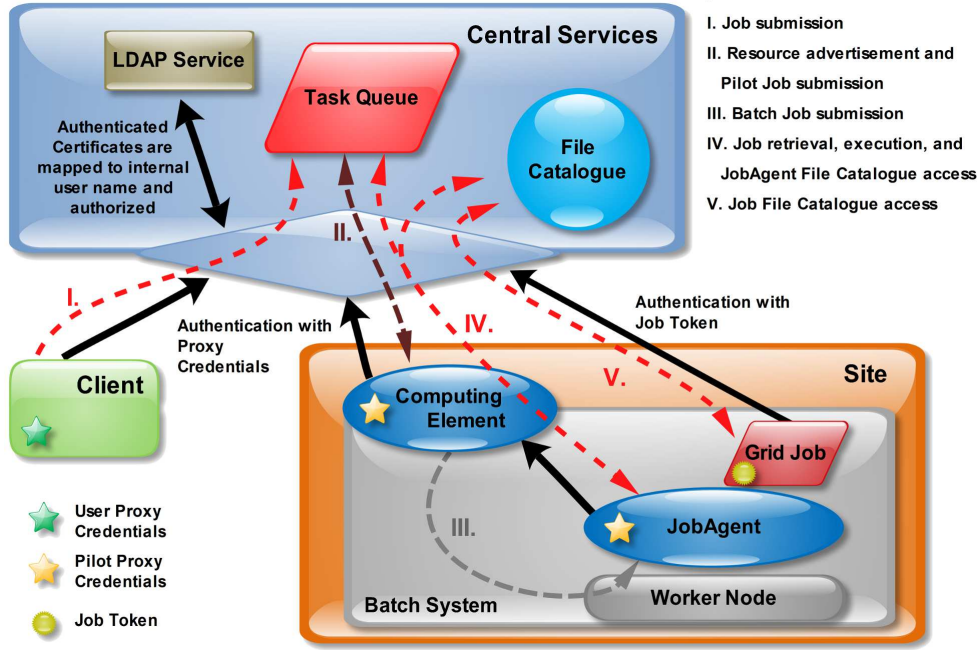}
\caption{\label{supj}Single-user Pilot Jobs in the \alice\ Grid Services}
\end{center}
\end{figure}
The JA uses a PC in order to authenticate and authorize itself to the
Central Services (see figure~\ref{supj}). This PC (hereafter
referred to as \textit{Pilot PC}) is submitted to a WN by the
Site's Computing Element in combination with the JA request, and is based on
the Grid certificate of a Site or Grid operator. Once placed on a Site's
Computing Element, it is renewed automatically by periodical requests to a
MyProxy~\cite{myproxy2} service. MyProxy is a credential management service
that holds long-term uploaded PCs and provides authorized entities with
derived PCs of lower order on demand~\cite{myproxy1}. In a typical usage
scenario, a delegator uploads a long-term PC, having a validity e.g.\ of one
month. A Grid service identifying itself with an explicitly authorized Grid
credential
then can request short-term delegated PCs, e.g.\ for 24 hours, from the MyProxy
service. The delegator can refresh the long-term PC as needed.
\\
As the JA executes Grid jobs directly, all jobs are executed on WNs
using the same local user account as their respective JA and Grid job user accounting is
based only on a job submitter's internal user name in \alice\. With respect to a
Site, both the JA and all jobs are executed by the person
identified by the Pilot PC, respectively the Site or Grid operator.
A JA's Pilot PC has escalated privileges to impersonate a user
that submitted a given job, and to retrieve input files from and upload
output files to the Grid File System in their name. 
The job itself can utilize a security token in order to access the Grid file
system. It is created on request in the Central Services during the
initialization of the job on a WN and revoked once the job finishes.\\
A JA does not affect a job's execution beyond
monitoring and potential termination due to expiration or resource limit 
exceedance. Both the 
PJ and its jobs are executed conforming to a Platform as a Service (PaaS)
model.
The operating system and the JA represent the environment of the Grid job.
While the administrative sovereignty of the WN resides at
the Site, the sovereignty of the JA program code is
at the VO.\\
\par
\begingroup
\leftskip=0.4cm
\noindent
\textbf{Definition 1.1}: 
The approach of multi-user Grid jobs executed by Pilot Jobs which run with
single VO-specific users on a WN is referred to as a model of
\textit{Single-User Pilot Jobs} (SUPJ).\\
\par
\endgroup\noindent
In contrast to Single-User Pilot Jobs, we define an according multi-user model
as follows:\\
\par
\begingroup
\leftskip=0.4cm
\noindent
\textbf{Definition 1.2}: 
The Execution of Grid jobs from different users in a Pilot-Job-based Grid
infrastructure with corresponding different local user accounts on a WN is
referred to as a model of \textit{Multi-User
Pilot Jobs} (MUPJ).\\
\par
\endgroup\noindent

\section{Security Analysis of Single-User Pilot Jobs in \alice}
\label{secsupj}
For the analysis of the accountability of Grid users as individuals,
the \alice\ VO is considered to act as a service provider for its users, and as an
intermediary between its users and the Sites as service and platform providers.
As this intermediary, it is known as a central broker entity to all Sites,
and decides which Grid jobs are supposed to be executed by which Site, and
consecutively submits the jobs to the Site provided WNs. 
Accordingly, the VO
receives \textit{task assignments} from its users, while propagating these
assignments as Grid jobs to the Sites. Along with the assignment, a Grid user
performs an implicit \textit{delegation} of a corresponding subset of its
privileges to the WN and the corresponding Site. The Grid Services in
combination with the \aliroot\ framework as a software package provide a
Software as a Service (SaaS) to its users, which is based on the PaaS level usage of
the Sites' resources.
Aside, the Grid Services provide a PaaS to the user's, as any code can be
supplied and requested to be executed as a Grid job. The
program code and data executed on a WN along with a Grid job in the \alice\ Grid
Services can be classified according to three internal and one external
origins, as specified in table \ref{dataorigins}.\\
The model of Single-User
Pilot Jobs, as implemented within the \alice\ Grid Services, has
drastic limitations to security and user accountability. In the following
paragraphs, these limitations are discussed and security objectives are defined
in order to allow for an analysis of potential solutions.
\begin{table}[h]
\begin{center}
\begin{tabular}{p{0.32\textwidth} p{0.625\textwidth}}
\hline
\noalign{\medskip}
Internal I: Detector and user data, and user code.
& Data stored in the Grid File System within files. This can be the data
of the \alice\ detector, both raw or preprocessed through several
stages, or any data or program code supplied by user's of the
system.\\
\noalign{\bigskip}
Internal II: Software Packages
& Program code downloaded and supplied to the job by
the JobAgent, provided by the VO as software packages.\\
\noalign{\bigskip}
Internal III: Worker Node
& Program code provided within the operating system of the WN, such as
system commands and libraries.\\
\noalign{\bigskip}
External
& Data retrieved within the job directly from external or third party
resources, e.g. via downloads from the Internet.\\
\noalign{\medskip}\hline
\end{tabular}
\caption{\label{dataorigins}Data origins in the \alice\ Grid Services}
\end{center}
\end{table}
Due to the current architecture of job submission and execution, it is
virtually impossible to state the origin of potential security incidents,
attacks, or misbehaviour arising along or from a Grid job executed on a WN.
Once a user has submitted a job to the \alice\ Grid Services, the job is
completely in the sovereignty of the VO. As the relation of a Grid user and a
job is provided by the internal user name only, this relation is fully
controlled within the Central Services. Similarly, the VO has the control to deliberately
alter a user's job submission. Within the Central
Services and before their submission to Sites, the original job requests are
processed. In the course of this processing, jobs can be split into sub-jobs
working on only subsets of the specified input data. Once at a Site, there is
no further insurance of a correct execution of a job beyond simple run time and
resource utilization monitoring by the Pilot Job. Finally, as a consequence, a
Grid user has no possibility to prove non-conformity of the actions taking place 
in their name and as such to protect the misuse of its identity, which results
in the following security objectives:\\
\par
\begingroup
\leftskip=0.4cm
\noindent
\textbf{Objective 1}, \textit{Provable authenticity of assignment}:
The original submission of a Grid job must be verifiable at any later stage,
proving the submitter's identity and the assignment as it was submitted.\\

\noindent
\textbf{Objective 2}, \textit{Provable authenticity of assignment processing}:
The processing of a Grid job as an assignment must be
verifiable at any later stage, proving to result from a set of sound
transformations performed within the Central Services, respectively by the VO
only.\\

\noindent
\textbf{Objective 3}, \textit{Protection against forgery of assignment}: 
Grid job assignments must not be forgeable by Grid users, the VO, the Sites, or
any third party.\\

\noindent
\textbf{Objective 4}, \textit{Protection against misuse of delegation}:
The delegation of privileges along with a Grid user's job submission must be
protected from being misused.\\
\par
\endgroup\noindent
The objectives 1 to 4 represent necessary criteria for non-repudiation of
both a Grid user's job submission and the job's processing within the VO's
control.\\
According to the SUPJ model, the JA on a WN executes all Grid jobs as
received on behalf of the \alice\ VO, in which the jobs are executed directly by the JA
process and therefore run within the same local user
environment on the WN. Hence, Grid jobs are not strictly isolated from each
other within their execution. A JA runs only one Job at a time and a job's
working directory is scratched after the execution. Nevertheless, a job can fork sub-processes
that will remain after its execution on the system, running with the same user
account and privileges, and are therefore for example able to alter later
executed jobs. Further, jobs are neither encapsulated nor isolated with respect to
their JobAgent, and are therefore able to alter the JobAgent or get hold of the
Pilot PC. This introduces a crucial security issue to the Grid. A Grid job could
start e.g. the \alien Grid client on the WN, while utilizing the Pilot PC, and
submit new jobs. This would enable an attacker to conduct for instance denial of service
attacks. Further on, the Pilot PC entitles to act on any user's behalf within the
\alice\ Grid, as this is needed in order to handle file uploads in the name
of the job and thereby in the name of the job submitter. As a consequence, any
holder of a Pilot PC is currently able to impersonate any \alice\ user. 
Accordingly, we further formulate the following security objectives:\\
\par
\begingroup
\leftskip=0.4cm
\noindent
\textbf{Objective 5}, \textit{Grid job isolation}: Grid jobs should be
mutually isolated and must be prevented from potential mutual interference,
both concurrently and consecutively in time.\\

\noindent
\textbf{Objective 6}, \textit{Pilot Job protection}: A Pilot Job must be protected from alteration, 
interference, or
disruption by one of the Grid jobs it is executing. Analogously, its Pilot Job
credentials need to be protected from any misuse by jobs.\\

\noindent
\textbf{Objective 7}, \textit{Pilot credential limitation}:
Pilot Job credentials must be limited in power, not to allow any escalated
privileges, in particular with regard to Grid user's identity, in order to 
impede misusage by Grid users, the
VO, the Site, or third parties
must be protected from any misuse by a
Grid user, the VO, a Site, or thirds.\\

\noindent
\textbf{Objective 8}, \textit{Pilot platform integrity}:
The WN and its Pilot Jobs as the Grid job platform, must provide an
environment of integrity and be secured of any non-conform Site access or
access of any third party.\\
\par
\endgroup\noindent
Grid jobs originating from different users are not visible to
a WN's operating system, and by that to the Sites, in a transparent way and it
is not possible to enforce a per-user Grid job control. In case of security incidents or
attacks originating from single Grid user accounts, it is not
possible to respond accordingly and potential counter-measures can only affect
a VO's entire set of jobs on a Site or WN. Finally, the architecture prepares
no possibility for Site-based usage and resource accounting on a per-user
level. Accordingly, we state the last further
objective:\\
\par
\begingroup
\leftskip=0.4cm
\noindent
\textbf{Objective 9}, \textit{On-Site Grid job user accounting}: Grid
jobs need to be authenticated and authorized in a transparent way in the
operating system of a WN.\\
\par
\endgroup\noindent

\section{Multi-User Pilot Jobs with gLExec and Proxy Credentials}
\label{mupj}
In order to allow for a secure handling of Multi-User Pilot Jobs, the Grid middleware
gLExec~\cite{glexec} was developed. Instead of a direct execution of a job or
payload, gLExec can be invoked by a PJ, in order to enable authentication and
authorization of an associated identity and to allow for isolation of a job's or
payload's process. The authentication and authorization is based on the proxy
credential of a Grid user, hereafter referred to as \textit{Grid user
Proxy Credential} (GuPC), provided to gLExec through an environment variable.
Isolation of a Grid job, and its separation from the PJ, can be obtained through a 
user and environment switch within a POSIX-compatible operating system, 
similar to the UNIX \texttt{sudo} command.
Depending on its configuration,
gLExec would map a given GuPC to a local POSIX user ID whose value would
usually be different for distinct values of the Grid Certificate's Subject line.
It is also possible for a site to configure gLExec without such an identity change,
whereby the user job or payload will run with the same local user ID as the PJ
itself (assuming the GuPC was authorized).
Within this
document, we will consider the utilization of gLExec only for the case of a
full invocation with enforced authentication and authorization and a subsequent
certificate-mapping-based user switch. Moreover, we
assume this operating mode to be able to comply with
objectives 5 and 6 defined above. This assumption implies a sound setup and
invocation of gLExec and lies within the limitations of the achievable degree
of isolation of 
different user accounts inside the same POSIX-compatible operating system.

\subsection{Security limitations of unrestricted Proxy Credentials}
\label{proxies}
The X.509 PCs as they are
in use throughout the \wlcg\ and beyond are based on unrestricted delegation.
As such they have long-known cardinal security limitations~\cite{proxy}, which
were already considered while X.509 PCs were being adopted as a functionality~\cite{x509delegation}.
In this section, we specify and describe from a 
conceptual perspective four
essential limitations of PCs
with respect to their usage
in the \wlcg\ (while disregarding auxiliary restrictions that can be 
applied beyond):\\ Unrestricted delegation based on proxy
signatures can be illustrated by a mathematical representation of
the delegation as a function.
Let
$U = \{ u: \textrm{user able to delegate privileges}\}$,
$P_{u} = \{ p_{u}: \textrm{delegable privilege of a user } u \in U\}$,
and $T = \{ t: \textrm{time stamp in seconds} \}$.
With  
$t_{issued},t_{expires} \in T$ as the two time stamps of beginning and
end of validity, the delegation of
privileges based on the usage of X.509 proxy certificates, can be expressed as the function
\begin{center}
\begin{equation}\tag{f 3.1}\label{e3.1}
\gamma_{\text{PC}}: U \times T \rightarrow
\{\bigcup \limits_{u \in U} P_{u},\emptyset\}, \;\;\ \text{with} \;\;\
\gamma_{\text{PC}}(u,t) = \begin{cases} \; \emptyset \;\;\, \quad \text{if}\;\;t
< t_{\text{issued}} \;\ \text{or} \;\ t \geq t_{\text{expires}} \\ \; P_{u}
\quad\;\text{if}\;\;t_{\text{issued}} \leq \; t\; <
t_{\text{expires}}\quad\quad.
\end{cases}
\end{equation}
\end{center}
As such, the delegation is not only unrestricted, but also has
no dependencies other than in the dimension of time.\\
\par
\begingroup
\leftskip=0.4cm
\noindent
\textbf{Limitation 1}, \textit{Unconditional delegation}:
A PC itself has neither any binding to a particular delegate nor any
context-sensitivity of its usage. Due to this, any privilege is held as
such and any limitation or binding would require additional external mechanisms in
place.\\

\noindent
\textbf{Limitation 2}, \textit{Unrestricted delegation}: 
Except in time, a PC allows only for an unrestricted delegation to the
delegate and thereby holds all privileges of the delegator as such.\\
\par
\endgroup\noindent
A PC within the \wlcg\ has a validity of typically hours or days, which
cannot be considered too little for a successful 
exploitation by potential attackers. In scenarios including the use of MyProxy
services, a PC can in principle be extended in
lifetime by a renewal request to a corresponding MyProxy service, as long as the
original first order PC inside the service is still valid. The renewal
can be controlled by the use of keys and usually is based on specific properties
of the requester's own PC (typically its Subject),
though the effectiveness of these mechanisms is
dominated by secrecy of the keys and security of the involved systems.\\
An extension to the proxy mechanism called VOMS (Virtual Organization
Membership Service)~\cite{voms} provides the possibility to apply authorization attributes
to a PC. On presentation of a valid PC to a VOMS server,
an entity can request the addition of any attributes to which it is entitled,
e.g. VO membership or roles. Security can be improved by application of a VOMS-based
authorization setup in which a plain PC without the necessary VOMS extensions
would not be sufficient to conduct any tasks in the system.
However, within the \wlcg\ environment
the VOMS extensions are primarily
used to elevate the privileges of a PC holder, e.g. to obtain the necessary role
to run a PJ. Without explicit preventions, any holder
of a plain PC can request a new set of VOMS extensions within the range of
privileges of the original certificate owner.\\
As such, PCs may be, once obtained and within their validity in time,
fully exploited for identity theft and their misuse could lead to severe security
consequences. Regarding GuPCs, this discussion can be further
substantiated by the \wlcg\ \egee\ Grid Security Policy: \textit{"Users
[\ldots] must ensure that others cannot use their credentials to masquerade as
them or usurp their access rights. Users may be held responsible for all actions
taken using their credentials, whether carried out personally or not. No intentional
sharing of credentials for Grid
purposes is
permitted."}\ifthenelse{\boolean{bpublic}}{~\cite{secpolano}}{~\cite{secpol}}\\

\par
\begingroup
\leftskip=0.4cm
\noindent
\textbf{Limitation 3}, \textit{Exposure to theft}: A PC is by itself
completely unprotected, while being handed on within a distributed system such
as a Grid environment. Regarding this aspect it is comparable to a plain
security token. Without additional protection a PC can be stolen at any point
where it resides and must be expected to be accessible
by persons with privileged access.\\

\noindent
\textbf{Limitation 4}, \textit{Multiple domain validity}:
A Grid Certificate and hence a PC can be recognized by more than one VO or
set of service providers, and thus be used to access all the resources
concerned.\\
\par
\endgroup\noindent
Limitation 4 would apply when different VOs or sets of
service providers accept the same PC for a particular person.
Within the \wlcg\ VOs such a situation is very rare: some persons are members
of multiple VOs, but usually with a unique Grid certificate Subject per VO
(for practical reasons).
The limitation would in particular affect users with
administrative functions in at least one of their VOs. As a consequence of the
limitations 3 and 4, a low level of protection of PCs in one VO
could drag down the level of protection in another VO.

\subsection{Propagation of Grid user Proxy Credentials}
\label{prop}
The application of gLExec relies on the propagation and transmission of GuPCs
from the actual job submitter and owner of a Grid Certificate to the Pilot on
the WN where the job is supposed to be executed.
Yet on the part of gLExec, there are no
specifications or particular information regarding a potential implementation of
this process. We therefore briefly outline two
different approaches taken by the \lhc\ experiments
\textit{\atlas} and \textit{\lhcb}:\\
In case of
\atlas\ifthenelse{\boolean{bpublic}}{~\cite{atlas1ano,atlas2ano}}{~\cite{atlas1,atlas2}},
the adoption of gLExec is based on the utilization of one central MyProxy service
located at \cern, into which GuPCs are uploaded protected with random keys that are
kept within the VO's sovereignty in their WMS implementation.  Once a PJ
on a WN connects to the WMS in order to retrieve a job,
together with the job description it will receive the key, which is then used by the PJ to
retrieve a valid user GuPC from the MyProxy service.\\
\par
\begingroup
\leftskip=0.4cm
\noindent
\textbf{Definition 3.1}: The approach in which a VO holds keys with a
one-to-one relation to the actual GuPCs, although it will not store or
transport the GuPC directly, is considered
an \textit{indirect GuPC propagation}.\\
\par
\endgroup\noindent
The integration of gLExec into the \lhcb
\ifthenelse{\boolean{bpublic}}{~\cite{lhcbano}}{~\cite{lhcb}} WMS follows a
different methodology, while not utilizing the MyProxy service for that purpose.
The GuPCs are managed
directly by the VO's central service and
the pilot receives the GuPC together with the payload.\\
\par
\begingroup
\leftskip=0.4cm
\noindent
\textbf{Definition 3.2}: The approach of a proprietary storage within the
VO's sovereignty and implicit transfer of GuPCs is considered a
\textit{direct GuPC propagation}.
\par
\endgroup\noindent

\subsection{Consequences of Grid user Pilot Credential propagation}
\label{problems}
Without comprehensive additional mechanisms, the above
described limitations of PCs based on unrestricted delegation
lead to fundamental
weaknesses concerning Grid job user accountability. Their adoption
as GuPCs can
introduce severe security
threats in matters of MUPJ frameworks. Subsequently,
these weaknesses and threats are discussed as security problems:\\
\par
\begingroup
\leftskip=0.4cm 
\noindent
\textbf{Problem 1}, \textit{Unprovable correlation of assignment and
delegation}: A job submitter's GuPC on a WN does not have any binding to any
actual job or payload. The availability or presence of a valid GuPC is neither a
binding statement to prove the authenticity of a job requested to be executed on
a WN, nor does it prove to be an authentic derivative of
a user's submission.\\
\par
\endgroup\noindent
As a consequence and without further controls, the use of GuPCs is not
able to fulfill the requirements
for accountability of users with respect to job submissions. GuPCs could be
potentially stolen, misused,
or mixed up at various points between the user's job submission
and a WN without notice. Similarly, 
a job's description or
payload could be altered or exchanged. In
\ifthenelse{\boolean{bpublic}}{\cite{egeeano}}{\cite{egee}} this concern was
raised as the necessity to trust a VO to provide flawless correlations between
GuPCs and submitted jobs.\\
\par
\begingroup
\leftskip=0.4cm
\noindent
\textbf{Problem 2}, \textit{Fuzzy validity and expiration}:
The validity of a GuPC is by itself independent of the validity or lifetime
of a Grid job.\\
\par
\endgroup\noindent
Without explicit functionality, a GuPC must be assumed to be
still valid once a corresponding Grid job has terminated. In case of an
indirect GuPC propagation using a MyProxy service, the relation
between GuPCs and Grid jobs cannot be assumed to be bijective, viz.\ the same
GuPC could (and actually will) be
used for several Grid jobs, to reduce the credential management overhead.
In any case, a GuPC could potentially be renewed, 
a GuPC could potentially be renewed until the corresponding
first-order GuPC stored in the MyProxy service expires. In principle also the
latter credential could even be renewed periodically, which
would lengthen the potential validity of all GuPCs derived from it. In the
worst case, this could lead to GuPC validities up to many months.\\
\par
\begingroup
\leftskip=0.4cm
\noindent
\textbf{Problem 3}, \textit{Unlimited access of VO and Sites}: Even if GuPCs
are never stored or processed within a VO WMS, as with
the indirect GuPC propagation, a user or attacker holding certain privileges
within the
WMS must still be considered able to retrieve any active
user's GuPC. Since a GuPC must be readable to the
PJ process on the WN which manages the handover to gLExec, at least anybody
who can control the PJ (within the VO or the Site) would be able to retrieve GuPCs.\\

\noindent
\textbf{Problem 4}, \textit{Elevated trust in the VO and Sites}: In
comparison to the SUPJ scenario, both direct and indirect GuPC propagation for MUPJs
signify an elevated level of necessary trust in the VO
and the Sites. While in the SUPJ any user
or attacker that has access to certain privileged services can only harm the concerned VO itself
and exploit a user's
privileges within the VO, the holder of a GuPC might also access a
third VO or service that happens to recognize the GuPC. As a consequence,
not only do the certificate owners have to put an increased trust in all VOs
they are working with, but also the required level of mutual trust between VOs
is raised accordingly.\\

\noindent
\textbf{Problem 5}, \textit{Challenge of storage}:
For both the direct and the indirect GuPC propagation the main storage
entity of GuPCs becomes a critical security concern. The storage must be
instantiated and maintained secure. Attacks on the storage must
be considered severe security threats.\\

\noindent
\textbf{Problem 6}, \textit{Drawback of additional service invocation}:
A scenario utilizing remote service callbacks, e.g. the indirect GuPC
propagation, introduces additional risks of mitigated
availability due to failures or attacks. Moreover, any additional 
invocation amounts to additional dependencies, additional load in matters
of scalability and the introduction of delays.\\
\par
\endgroup\noindent
As a consequence, the application of gLExec based on the presented
alternatives for the implementation of the propagation
of GuPCs amounts to no change or improvement regarding the significance
in accountability. In both
the SUPJ and the MUPJ scenarios described above, the VO is able to submit jobs
to Sites in the name of a user with neither the user's nor the Site's notice.
Consequently, the presence of a user GuPC cannot be considered at all as
proof or anchor for accountability and an implementation would not be able to
fulfill the criteria reflected by the objectives 1 to 4, 7, and 9.
A sound implementation of job accountability does not only
allow for a user to be proven accountable for a certain
malicious or illegal behaviour that was observed, but also ensures a user can
rightfully
disclaim responsibility when the user's behavior was appropriate. Consequently,
the objective must be to allow not only for identification of the user who
submitted a job, but also for verification of the actual job at hand.
In order to prove the authenticity of a Grid job on a WN, first its origin
needs to be provable to be an authentic submission by a particular user
and second, any alterations of the job by the VO need to be verifiable.

\section{Mediated definite delegation}
\label{model}
By analyzing the job submission model in the \alice\ Grid Services, we found no
reason to justify the utilization of a full delegation approach in
order to allow for MUPJ. In contrast, JDL-based job requests
describe the required permissions in a context-sensitive
expression, and as such implicitly state the least privilege of a necessary
delegation.\\
Facing a three-tier design of service users (Grid users), a service
broker and processor (VO), and a service/platform provider in the back end (Sites), 
the delegation by a user to a Site or WN in the course of a Grid job is mediated
by the VO. Aligned to this design, we present a new mediated delegation
model based on the delegation of definite privileges, in order to access explicitly
specified system entities, to agents assigned by a broker after applying
verifiable transformations.\\
With respect to Grid environments, let
$U = \{ u: \textrm{user able to delegate tasks}\}$,
$P = \{ p: \textrm{delegable privilege}\}$,
$E = \{ e: \textrm{accessible entity, as e.g. files, jobs, or
services}\}$, $A = \{ a: \textrm{agent, able to execute tasks
on behalf of users}\}$, and $T = \{ t: \textrm{time stamp in seconds} \}$.
Further let $C = \{ c: \textrm{concession}\}$, wherein each element describes
the delegation of $p \in P$ in the name of $u \in U$ with respect to $e \in E$ to an $a \in A$ 
at a certain point in time $t \in T$.\\
\par
\begingroup
\leftskip=0.4cm 
\noindent
\textbf{Definition 4.1}: A
delegation based on the transfer of concessions is defined as \textit{delegation of tasks}
or \textit{static delegation}, and is expressed by the following mapping
\par
\endgroup\noindent
\begin{center}
\begin{equation}\tag{f 4.1}\label{e4.1}\delta: U\times P \times E \times A
\times T \rightarrow C \;\;.\end{equation}
\end{center}
Let $B = \{ b: \textrm{task request broker}\}$, $ \bar{C} =
\{ \bar{c}: \textrm{non-mediated concession} \}$ describing delegations of $p \in P$ in
the name of $u \in U$ with respect to $e \in E$ at a certain point in time $t \in T$
to be mediated by a $ b \in B$, and 
$D = \{ d: \textrm{derivative, a verifiable transformation}\}$ 
describing derivatives a $b \in B$ can apply to a $\bar{c} \in \bar{C}$ upon mediation.\\
\par
\begingroup
\leftskip=0.4cm 
\noindent
\textbf{Definition 4.2}: $\bar{C}$ is defined by 
a \textit{mediation request for task delegation} as a mapping
\par
\endgroup\noindent
\begin{center}
\begin{equation}\tag{f 4.2}\label{phi}\phi: U \times P \times E \times B \times
T \rightarrow
\bar{C}\;\;.\end{equation}
\end{center}
\par
\begingroup
\leftskip=0.4cm 
\noindent
\textbf{Definition 4.3}: A \textit{mediation of task delegation} is defined as
the derivative or transformation of a $\bar{c} \in \bar{C}$ according to a $D$
and the assignment to an $ a \in A$ at a certain point in time $t \in T$,
with the result being an element of $ \hat{C} =
\{ \hat{c}: \textrm{mediated concession} \}$. It is expressed by the mapping
\par
\endgroup\noindent
\begin{center}
\begin{equation}\tag{f 4.3}\label{psi}\psi: \bar{C}
\times \mathcal{P}(D) \times A \times T \rightarrow \hat{C}\;\;.\end{equation}
\end{center}
A mediated delegation of tasks or \textit{mediated definite delegation}
can then be expressed by a composition of the mappings, as \\\\
\centerline{$\delta_{\text{mediated}}(u, p, e, b, t', D, a,
t'') = \psi(\;\phi(u,p,e,b,t')\;,D,a,t'')\;\;.$}\\\\
The mapping $\delta_{\text{mediated}}$ then describes a definite static
delegation with respect to the entities the delegated privileges apply to, but is dynamic with
respect to the agent to be elected and using the delegated privileges during
task execution.\\
Within the \alice\ Grid Services, the Central Services act as a broker,
deciding where a Grid job is to be executed and thus determining which
agent receives a Grid job for execution. The JDL of a job is independent
of this decision, and any transformations in the course of its processing within the Central Services 
describe a refinement or derivative of the specified information. As such, it is
possible to view the processing as refinements, while only appending
information to the original JDL.\\
The \alice\ Grid File System is based on the abstraction of a
logical and a physical layer, with a File Catalogue containing
logical file entries as references to physical files stored in
distributed storage services. By replicating logical files to several 
storage servers the system is
improved with respect to resilience and allows for optimizing data access
according to
proximity\ifthenelse{\boolean{bpublic}}{~\cite{sediscano}}{~\cite{sedisc}}. The
File Catalogue access from within Grid jobs is based on this mechanism, and the 
JDL refers to logical file entries,
while allowing the user to specify storage servers as preferences or
constraints. The access is granted based on a central ticketing service, which 
incorporates control over the storage server selection. Accordingly, the
physical file access requests in line with a Grid job can be modelled as
derivatives of the logical specification within a JDL.\\
Finally, the model can be extended to allow the propagation of
non-mediated concessions between several brokers before their assignment to an agent.
Therefore, the set $\bar{C}$ is
redefined, so its elements can be assigned not only to an $a \in A$  but as
well to a $b \in B$. Then a function $\varrho$ must be defined to
allow for a\\\\
\centerline{$\delta\rm{^n_{mediated}} =
\psi(\;\varrho^n(\;\phi(\ldots)\;,\ldots)\;,\ldots)$\;\;,}\\\\
where $n$ is the
number of subsequent applications of the mapping
$\varrho$ representing relaying between brokers. The case $n=0$ represents the
mapping $\delta_{\text{mediated}}$  as defined above.\\
\par
\begingroup
\leftskip=0.4cm 
\noindent
\textbf{Definition 4.4}: A \textit{relaying of mediation request} is defined as
the propagation of a $\bar{c} \in \bar{C}$ to another $ b \in B$ while applying
derivatives or transformations according to a set $D$ at a certain point in time
$t \in T$. It is expressed by the mapping
\par
\endgroup\noindent
\begin{center}
\begin{equation}\tag{f 4.4}\label{relay}\varrho: \bar{C}
\times \mathcal{P}(D) \times B \times T \rightarrow \bar{C}\;\;.\end{equation}
\end{center}
 
\subsection{Certified Grid jobs}
\label{sjdl}
The described model of mediated delegation can be implemented based on 
composite digital signatures of a JDL in the job request.
As user Grid Certificates together with their corresponding private keys can
be used for digital signatures, we
propose to require a client to sign the JDL upon submission.
By utilizing a signed JDL (hereafter referred to as an \textit{sJDL})
throughout the whole propagation of the job, while also providing the corresponding 
user Grid Certificate, it is 
possible to prove at any point the integrity of a job and to identify illegal
modifications. A signature allows proving the authenticity of the
JDL using the user's public key, while the Grid Certificate states a user's
membership and authorization within the VO.
The user Grid Certificate is appended for both signature verification and
certificate validity evaluation. Using time stamps within the sJDL, set by the user and
verified by the VO upon submission, an sJDL allows for a delegation with
warrant, stating the user wanted a particular job to be executed as such and within
the specified time frame.\\
In order to further
state the approval of a job to a Site, a central service of the VO signs the
sJDL as well, viz.\ as a broker, using an appropriate public-private key pair of a
VO-specific Grid Certificate. This second
signature is applied before a job is sent to a Site and includes all
information appended by the VO and states potential job transitions,
e.g. splitting. This doubly signed JDL will hereafter be referred to as
\textit{s$_2$JDL}. On the WN an authenticated JA gets authorized
as a delegate of the job it is asked to execute,
according to statements in the job's
s$_2$JDL. For (long-term) accountability the 
 s$_2$JDL needs to be stored within the Central
Services. If additionally recorded at the Site, it can be used as proof
of the retrieval of a particular job.
\begin{figure}[h]
\centering
\includegraphics[width=\textwidth,keepaspectratio]{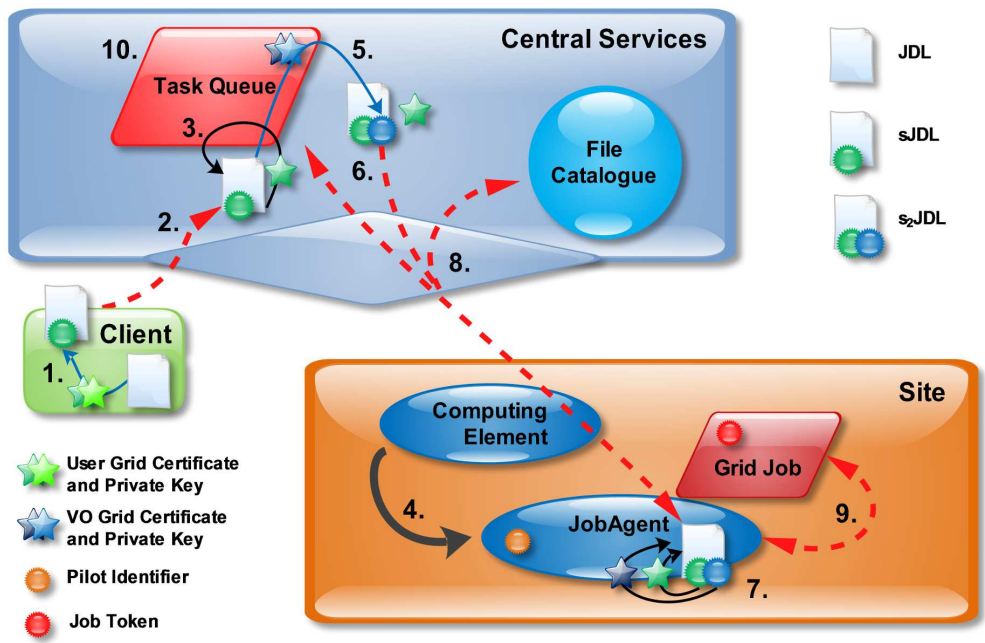}
\caption{Certified Grid Jobs}
\label{sJDL}
\end{figure}
The proposed protocol for Certified Grid Jobs in \alice\ is
detailed in the steps below and in figure \ref{sJDL}. 
The Client and Computing Element authentications based on PCs stay
unchanged and are omitted.\\
\begin{enumerate}[{Step} 1{:}]
\item The client sets a submission and expiration time stamp in the JDL and
signs it using the user's private key.
\item This sJDL is sent to the central service of
the VO together with the user's Grid Certificate.
\item The signature of the sJDL and the submission and expiration time stamps
are validated within the Central Services.
\item Upon request from the Central Services
the Computing Element submits a JA as a batch job,
with a JA identifier and the certificate identifying the Central Services.
\item The JA authenticates itself to the Computing Element with its
identifier and requests a job from the Central Services through the Computing
Element. In case of a matching job waiting in the Task Queue, the Central
Services sign the sJDL, including a submission and expiration time stamp
and the JA identifier. The use of a second pair of time stamps is
necessary as a job could be resubmitted by the VO in case of errors.
While the user's submission and expiration time stamps define the
time window during which
the VO is entitled to send the job to Sites, the VO's time stamps
in the second signature can be adapted to the desired run time of a job on a
worker node and as such can define a much shorter time window. 
\item The resulting s$_2$JDL is sent together with the corresponding user's Grid
Certificate to the JA and the job
is marked in the central Task Queue to be taken by this JA.  
\item The JA verifies both signatures, based on the authenticity of the
signatures, the validity of the time windows of submission and
expiration time, and the two certificate chains. In order to record a job's 
execution for accountability, the s$_2$JDL can be logged within the Site.
\item Although there are currently no limitations on read access 
in case of the \alice\ Grid services, the s$_2$JDL could be used for 
explicit restriction. Once a job has finished or is stopped, the JA
uploads the job's specified output files in the name of the job submitter,
while receiving
only write permissions as stated in the s$_2$JDL on file or directory level.
\item The Grid File System write access for the job itself can be
limited according to the s$_2$JDL as well, if users are required to specify all
output files explicitly in advance. Before a job is executed, the PJ could start
a communication
server which offers a local service in order to
provide Grid File System access for the \aliroot\ framework within a job.
The connection would be secured by a key which is propagated to the job upon
its start-up.
\item Once the output files of the job are uploaded by the PJ,
the job reaches a final state (e.g. \textit{DONE} or
\textit{ERROR}) and the use of the s$_2$JDL as a token is invalidated in the
central services.
\end{enumerate}

If the PJ directly or indirectly manages all external communications,
the job itself no longer needs a
connection to the Grid services. Thereby, all job-related accesses on
the Grid File System can be authenticated by the JA's identifier and authorized
based on the s$_2$JDL.
This design directly allows for a future adoption of gLExec.\\
The composite signature of the JDL represents the realization of the
\textit{mediated definite delegation} as formulated above in the
$\delta_{\text{mediated}}$ mapping. The first signature represents the $\phi$
mapping~\eqref{phi} with its sJDL outcome describing a series of
\textit{non-mediated concessions} as
elements of $\bar{C}$. Correspondingly, the second signature represents the $\psi$
mapping~\eqref{psi}, while its s$_2$JDL outcome
describes a series of \textit{mediated concessions} as elements of $\hat{C}$.
Furthermore, the implementation prepares for signatures of multiple brokers, 
according to a potential use case of the relaying of mediation requests~\eqref{relay}.\\
The mechanism allows for fulfillment of the security objectives 1-4, as
both the assignment's and assignment processing's
authenticities are provable and their forgery is prevented by the secrecy of the 
respective private keys. Regarding objective 7, the PJ identifier is only used
for authentication and awards no further permissions beyond
retrieving s$_2$JDLs in order to run the corresponding jobs. Therefore,
threats like the submission of new Grid jobs exploiting arbitrarily the identity of other users
within a job or any other attack or illegal behaviour
based on escalated privileges obtained through a Pilot PC are dissolved
(see section \ref{ppi}). Nevertheless, access to the PJ by a job and mutual job
interference cannot be impeded (objectives 5 and 6).
Regarding objective 9, \textit{On-site Grid job user accounting}, the
described scenario allows for logging of the certificate information of the
job submitter, which is provided along with the s$_2$JDL on the PJ's
request for a user job. However, Sites still have to
entrust the VO with the authentication and authorization of jobs, since the jobs'
verifications would take place inside the VO-supplied PJ code.\\
The functionality introduced for certified Grid jobs precisely follows
the current communication and service invocation schema of the \alice\ Grid Services
and thereby introduces no
additional remote invocations or callbacks between Grid job submission and
execution. It further requires no renewal of credentials or delegations.
The additional cost in computation due to
signature generations and verifications and the necessary storage can be
considered negligible. In line with a feasibility study of the mechanism proposed above,
a prototype
using an encrypted connection between the PJ and the Central Services was
implemented. The implementation is
based on standard security libraries only, utilizing
\textit{SHA384withRSA}-based
signatures provided by the BouncyCastleProvider\cite{bouncy} and encrypted
communication based on \textit{SSLSocket} from the \textit{javax.net.ssl}
package. A sample of an s$_2$JDL as utilized within the prototype and allowing for
serialization is given by listing \ref{lstsjdl}.
\begin{lstlisting}[captionpos=b,caption=A sample s$_2$JDL (identifier
and signatures truncated),
label=lstsjdl,basicstyle=\ttfamily\fontsize{7}{11}\selectfont]
<SJDL><NOTBEFORE>1312392035</NOTBEFORE><NOTAFTER>1313601635</NOTAFTER><NESTEDJDL>
  <SJDL><NOTBEFORE>1312392035</NOTBEFORE><NOTAFTER>1313601635</NOTAFTER><NESTEDJDL>
    Executable = {"cat"};
    Arguments = {"myInputFile"};
    InputFile = {"/catalogue/data/myInputFile"};
    Output = {"stdout","stderr"};
    User = {"testuser"};
    Broker = {"myVO"};
    HashOrd = "Executable-Arguments-InputFile-Output-User-Broker";
  </NESTEDJDL><SIGNATURE>FTi2ATSgQ[...]CoA0TG==</SIGNATURE></SJDL>
  PilotIdentifier = {"FpK0bE9P[...]Jq1zNx"};
  HashOrd = "SJDL-PilotIdentifier";
</NESTEDJDL><SIGNATURE>EMQlV0Wzg[...]r47ivk=</SIGNATURE></SJDL>
\end{lstlisting}

\subsection{Certified Grid Jobs and gLExec}
\label{sjdlgl}
Assuming a gLExec modification to allow for authentication and authorization of
s$_2$JDLs (largely an imitation of what the PJ does itself),
further fundamental improvements could be achieved: by validating the
s$_2$JDL and
the accompanying public user certificate, not only the submission by a certain user
can be ascertained (as in a GuPC-based scenario), but actually the submission of
the particular job at hand. Presuming the gLExec user-switching mode, \textit{Grid
job Isolation} and \textit{Pilot Job protection} (objectives 5 and 6) can be
fulfilled completely within the limitations of the operating
system's user separation, as well as \textit{Pilot credential protection} (objective 7).
Taking this further, Ref.~\ifthenelse{\boolean{bpublic}}{\cite{chepano}}{\cite{chep}}
describes a mechanism for trustable meta information on files in the File Catalogue to be
provided by the underlying storage systems. In combination with s$_2$JDL and
gLExec it would then be possible to ensure not only a job's authenticity, but also the
authenticity of the files referred to within its s$_2$JDL. Given the Pilot
protection and the mutual job isolation, a job cannot influence executable,
arguments, or input files of other jobs running on the same WN,
as these are controlled by the protected PJ and based on the sJDL.
Assuming \textit{Pilot platform integrity} (objective 8) including a sound
gLExec processing, this allows for full \textit{On-site Grid job
user accounting} (objective 9): the identities of files added by any user
would be recorded by the trusted storage systems in checksums, which are stored in
the File Catalogue.  A signed Grid job then acknowledges the user's intent for the
job to be executed as stated by the sJDL and thereby based on the referred files
as they appear in the File Catalogue at the moment of the job's submission. The time
stamps in the sJDL, the file alteration time stamps for
entries within the File Catalogue, and the checksum verifications during
transmissions of the files allow for detailed assessment and accounting of the
job before and after run time (see section \ref{nrepud}).\\
Our approach of s$_2$JDL-based job authentication and
authorization was presented to the gLExec development team and
has been accepted as a potential development in the form of a suitable plug-in.

\subsection{Non-repudiation of Certified Grid Jobs with
gLExec}
\label{nrepud}
In this section we discuss the conditions for accountability and
in particular 
non-repudiation with respect to an attack having arisen
from a certified Grid job executed via a modified
implementation of gLExec as previously specified. 
If not stated
otherwise, we will consider the Pilot platform to
perform with integrity (objective 8), as well as the PJ itself and the VO,
and the VO-provided software packages to be non-malicious.
We examine exculpatory and inculpatory evidence, while 
referring to the four different data origins specified in table
\ref{dataorigins}.\\
In case a hostile Grid job was executed,
malicious code is either provided directly in the input files within 
the Grid File System beforehand, or
retrieved from third party sources during run time. As part of the forensics
after discovery of the job, evidence may be found within the job's input
files, showing either malicious data or a request to retrieve code
from external sources.
A user could delete or overwrite the
files in the Grid File System to cover up traces. Nevertheless, with
respect to the job's run time, the file alteration time stamp in the File
Catalogue will record such changes.
Since the File Catalogue has the property that physical files to which
it refers are never overwritten or deleted directly, there exists a time window
for recovering the original data of an affected file entry.
Both delete and overwrite
operations on the File Catalogue level lead to a simple disassociation of the
affected physical files and shadow entries are stored in a central bookkeeping
table\ifthenelse{\boolean{bpublic}}{\cite{chepano}}{~\cite{chep}}. The success
of the recovery is limited by the time frame for deletion of stale files.\\
Conversely, a user would be able to disclaim responsibility for
malicious code execution within a job when that job only referred to
files uploaded to the Grid File System by other users.
In that case the challenge is to find the evidence in those files.\\
In case of malicious software packages or an infected or attacked batch system
or WN, a disclaimer of responsibility would be more difficult to assess.
Nevertheless, the proposed
mechanisms could ensure no counterfeit evidence for the job submitter's
responsibility could be put in place.\\
If only storage systems are infected or attacked, the reference checksums in the
File Catalogue will identify modified data, as the checksums are verified after
the data has been received by the PJ or the job.\\
Finally, in case the File Catalogue is infected or attacked, it is possible to
place counterfeit evidence claiming a job submitter's responsibility for a malicious
job. This could be achieved due to file ownership, checksums, and modification
timestamps being stored in the File Catalogue and the sJDL relying on these
references. As such, the File Catalogue constitutes an entity which needs to
be fully trusted by users and Sites. The same applies to the ticket service
granting access to storage servers on the physical storage layer. If the PJ or
the job receives illegimate tickets, not conforming to the
specifications within the sJDL, accountability and verifiability are exposed.

\subsection{Pilot platform integrity}
\label{ppi}
Objective 8, the \textit{Pilot platform integrity}, cannot be assured by
the presented mechanisms and was introduced
as an auxiliary criterion. Nevertheless, the presented approach
of least-privilege Pilot Job credentials
and Certified Grid jobs is able to largely simplify the assessment of illegal or 
improper behaviour in a Grid job's environment on a WN.
By preventing Grid job submission with credentials stolen from a WN
the mechanisms would impede the proliferation of attacks.  Covering tracks
through escalated privileges and masquerading would be impeded as well.
Moreover, Grid
user credentials cannot be compromised and user identities can be protected
from misuse. Potentially exploited,
malicious file uploads to the Grid File System can be detected and cleaned up,
as the write privileges granted to a job are stated in its sJDL.

\section{Related Work}
\label{related}
Beyond the fulfillment of the defined
security objectives 1-9, a basic functional concern and boundary condition
was to identify approaches allowing the least invasive integration into the current architecture of the \alice\ Grid
Services, by implication not involving any additional remote callbacks or service
invocations (see problem 6). Accordingly, for example
GridShib-based~\cite{gridshib2} implementations
or dynamic restricted delegation~\cite{ondemanddeleg}, both based on callback mechanisms, were
disqualified. In Ref.~\ifthenelse{\boolean{bpublic}}{\cite{egeeano}}{\cite{egee}}
the signature of job requests is suggested as an approach to avoid a
potential ``mixup'' of GuPCs within a VO.\\
Snelling et al.~\cite{etd} proposed a model called \textit{Explicit
Trust Delegation} (ETD) to digitally sign job requests in the UNICORE Grid framework
allowing for static delegation. In comparison to our work,
ETD uses only one signature, either by the user or a trusted Grid portal, which
in the latter case is consequently based on unrestricted delegation to the
portal. Further, ETD does not distinguish entities such as broker and agent
within the Grid framework, and gives no explicit information on intermediate
processing, validation or the delegation's consequences for accountability
on the execution endpoint. 

\section{Conclusion}
This document examines security aspects of multi-user Grid jobs and
underlines, while referring to potential implementations, crucial 
deficiencies of unrestricted delegation based on X.509 proxy
credentials. Inspired by static delegation with warrant,
a new model of mediated definite delegation is presented,
allowing for dynamical assignment of
definite privileges of delegators to agents by task delegations.
Its implementation, based on compound digital signatures,
establishes verifiable task and privilege delegation statements,
thereby allowing for strong
accountability and long-term traceability of Grid job submissions and
establishing the foundation necessary for non-repudiation. The presented
prototype further impedes the escalation of privileges and prevents
identity theft on a WN. Its design foresees a potential interaction with
the gLExec Grid middleware and constitutes a necessary framework to
achieve full on-site user accounting and protection of Grid jobs and
their environment.

\section{Future Work}
The presented prototype implementation of Certified Grid
Jobs is part of a major
revision of the \alien\ Grid middleware, which will be further developed and
tested within the next months, featuring a completely new security
architecture. Within that process several aspects of the current discussion
will be followed up.  In particular: the formalization of the derivatives or
transformation rules for the job requests; the application of access
permissions on the physical storage level according to the access
granted on the Grid File System; further improvement of file
integrity assurance. We aim to incorporate in our
delegation model the aspect of Grid file access both on the logical
and on the physical layer and to provide for explicit file integrity assurance.\\
Finally, we will analyse and research the scope of security and
protection of the Grid job environment on a WN (objective 8).

\ifthenelse{\boolean{bpublic}}{}{
\section{Acknowledgment}
We would like to thank the gLExec development group for their valuable input and
suggestions and their positive feedback. In addition, we would like to thank
Olga Vladimirovna Datskova and Arsen Hayrapetyan for proofreading and their
efforts, help and suggestions regarding the formal presentation and
illustration of this work.
}

\bibliographystyle{abbrv}

\section*{References}
\bibliography{SignedJobs}

\end{document}